\documentclass[aps,prd,onecolumn,groupedaddress,showpacs,nofootinbib,amssymb
]{revtex4}
\usepackage[dvips]{graphicx}
\usepackage{amssymb}
\usepackage{amsmath}
\usepackage{epstopdf}
\usepackage{graphicx,,color}
\usepackage{amsfonts}
\usepackage{bm}
\usepackage{cancel}
\usepackage{comment}
\allowdisplaybreaks[4]

\begin{document}

\tolerance=5000

\title{Interplay between Swampland and Bayesian Machine Learning in constraining cosmological models}

\author{
Emilio Elizalde$^{1,2}$\thanks{E-mail: elizalde@ieec.uab.es} and 
Martiros Khurshudyan$^{1,2}$\thanks{Email: khurshudyan@yandex.ru, khurshudyan@tusur.ru}}

\affiliation{
$^1$ Consejo Superior de Investigaciones Cient\'{\i}ficas, ICE/CSIC-IEEC,
Campus UAB, Carrer de Can Magrans s/n, 08193 Bellaterra (Barcelona) Spain \\
$^{2}$ International Laboratory for Theoretical Cosmology, Tomsk State University of Control Systems 
and Radioelectronics (TUSUR), 634050 Tomsk, Russia \\}

\begin{abstract}
Constraints on a dark energy dominated Universe are obtained from an interplay between Bayesian Machine Learning and string Swampland criteria. The approach here differs from previous studies, since in the generative process Swampland criteria are used and, only later, the results of the fit are validated, by using observational data-sets. A generative process based Bayesian Learning approach is applied to two models and the results are validated by means of available $H(z)$ data. For the first model,  a parametrization of the Hubble constant is considered and, for the second, a parametrization of the deceleration parameter. This study is motivated by a recent work, where constraints on string Swampland criteria have been obtained from a Gaussian Process and $H(z)$ data. However, the results obtained here are fully independent of the observational data and allow to estimate how the high-redshift behavior of the Universe will affect the low-redshift one. Moreover, both parameterizations in the generative process, for the Hubble and for the deceleration parameters, are independent of the dark energy model. The outcome, both data- and dark energy model-independent, may highlight, in the future,  the borders of the Swampland for the low-redshift Universe and help to develop new string-theory motivated dark-energy models. The string Swampland criteria considered might be in tension with recent observations indicating that  phantom dark energy cannot be in the Swampland. Finally,  a spontaneous sign switch in the dark energy equation of state parameter is observed when the field traverses are in the $z\in[0,5]$ redshift range, a remarkable phenomenon requiring further analysis.

%\noindent{\small Research Areas: Alternative gravity theories; Dark energy; Evolution of the Universe.}

\end{abstract}

%\pacs {}

\maketitle

\section{Introduction}\label{sec:INT}

It seems clear that General Relativity cannot be the ultimate theory of the Universe, since it is unable to deal with situations which need to involve quantum physics (or even, maybe, an unknown theory beyond that one). The problem already appears explicitly in various viable modified theories of gravity, where quantum corrections turn out to be quite important~\cite{Nojiri}~-~\cite{Cognola}. One way out is assuming the existence of a well-motivated (but yet unknown) high-energy UV-complete theory, and also that General Relativity is the low-energy limit of such theory. Thus, it might be that we have been able to originate a working effective field theory,  from the low-energy limit of the true theory, which efficiently captures the behavior of the inflaton field and dark energy phenomena. Thus, we could be on the right track. In this regard, string theory may perfectly be a candidate for such UV-complete theory. It is  known that it has the capacity to unify the standard model of particle physics with gravity, but up to now the task of constructing dS vacua has hit a stonewall. Specifically,  no dS vacuum could be constructed yet, this leading to the assumption  that, in a consistent quantum theory of gravity, the dS stage does not exist~\cite{Kachru}~-~\cite{Danielsson}. This problem with dS vacua can indicate that they belong to the Swampland - the region wherein inconsistent semi-classical effective field theories inhabit\footnote{One starts from the assumption  that the whole landscape of vacua provided by string theory lead to consistent effective field theories~(EFT). And then, the problem with dS vacua, together with the fact that in the string landscape it is quite easy to obtain Minkowski and Anti-de Sitter solutions, leads to the conclusion that dS vacua do belong to the Swampland.} Thus, we might have a set of consistently-looking effective quantum field theories coupled to gravity, but they could turn out to be actually inconsistent with a quantum theory of gravity. This could again indicate that dS vacua are in the Swampland~\cite{Ooguri}~-~\cite{Obied}. 

Recently, various papers investigating the cosmological implications of two of the proposed Swampland criteria have appeared\footnote{It should be mentioned that in the recent literature there has been a lively discussion on the Swampland criteria from various perspectives, some of which can be found in~\cite{SCStart}~-~\cite{SCEnd}, while some of relatively old discussions can be followed in the list of references in~\cite{SCGP}}. In particular, some works have considered Swampland criteria under the form:
\begin{enumerate}
\item $SC1$: The scalar field net excursion in reduced Planck units should satisfy the bound~\cite{Ooguri}
\begin{equation}\label{eq:SC1}
\frac{|\Delta \phi|}{M_{P}} < \Delta \sim O(1),
\end{equation}
\item $SC2$: The gradient of the scalar field potential is bounded by~\cite{Obied}
\begin{equation}\label{eq:SC2}
M_{P}\frac{|V^{\prime}|}{V} > c \sim O(1).
\end{equation}
\end{enumerate}  
Provided we consider GR with the standard matter fields in the presence of a quintessence field $\phi$ to be the effective field theory\footnote{The two Swampland criteria  above demand that the field traverses a larger distance, in order to have the domain of validity of the effective field theory and $CS2$ to be fulfilled}. Here $\Delta$ and $c$ are positive constants of order one, the prime denotes derivative with respect to the scalar field $\phi$, and $M_{P} = 1/\sqrt{8\pi G}$ is the reduced Planck mass. A study of this kind is important in order to understand when the effective field theory, admitting solutions modeling an accelerated Universe, will not end up in the Swampland. One of the most recent analysis shows that the most accurate observational data hint towards the fact that the Swampland criteria may not be valid for a dark energy dominated Universe. The result has been obtained using a Gaussian Process and recently available expansion rate data. It should be mentioned that in the performed analysis, the authors did not use any dark energy model or a specific scalar field potential to estimate the bounds on the Swampland criteria. This owes to the fact that, applying a Gaussian Process, it is indeed possible to reconstruct the expansion rate and its high-order derivatives directly from the data, and then use them into the subsequent estimation process. On the other hand, it should be mentioned that the combination of reconstruction and estimation shows also that we could well start from the Swampland and end up in the Swampland. But, under some conditions, we might start from the Swampland and do not end up in it, eventually. And also the other possibilities exist: not to start in the Swampland and end up either inside  or still outside of it~\cite{SCGP}\footnote{This comes from the accepted interpretation of the results from the Gaussian Process}. 

The considerations above are an indication that an extra effort is still required to understand the ways how inconsistent theories should be separated from the consistent ones. Moreover, due to observational data, and most probably with the new higher redshift data, the situation will change and new models of a dark energy dominated Universe will be  developed. New interpretations for the dark energy will probably appearm too. This is therefore an interesting research topic; however, the goal of this paper is somewhat different. We would here like to establish a framework allowing to constrain cosmological models by using the string Swampland criteria, Eqs.~(\ref{eq:SC1}) and~(\ref{eq:SC2}), in the sense of the model being or not in the Swampland, and the same model ending or not in the Swampland, either. Probably, a study of this kind could not be  performed some time ago, but the situation now is different, as we shall see. Moreover, it should be mentioned that it is still not possible to perform this by means of Bayesian analysis, where  the likelihood function has to be evaluated. What renders the analysis now possible are the significant developments in Machine Learning \cite{PyMC3_p1}, \cite{PyMC3_p2}~(and references therein) occurred recently. In particular, when we consider the Bayesian Machine Learning approach that uses a model-based generative process, it becomes possible to perform the task of constraining the cosmological model by involving the string Swampland criteria presented above. Still, the constraints will need to be validated, in order to claim that the method works. In this regard, we can use one of the available observational datasets, in the way it will be shown in this work. Which could be namely a starting point to reconstruct new relevant dark energy models, inspired by or directly coming from string theory. This could also provide a reliable starting point to resolve the possible tension between the string Swampland criteria and the available observational data. 

More specifically, in this paper we will use Bayesian Machine Learning and string Swampland criteria, Eqs.~(\ref{eq:SC1}) and~(\ref{eq:SC2}) for  constraining cosmological models. The results of the fit are then validated using the best  expansion rate data available. Two simple models will be considered, one involving a parametrization of the Hubble parameter, and the other a corresponding one of the deceleration parameter. In both cases, the results of the fit from the Swampland criteria are then compared with the ones obtained from the generated expansion rate based on each model. Moreover, a generative process is used, with the advantage that then there are practically no restrictions on the redshift range, and also we are not limited by the specific characteristics and details of the observations. Even then, we are still able to involve information incorporated into the higher-redshift evolution of the Universe and, based on it, to improve the theoretical models. And this can also be used for forecasting criteria that will allow to improve and optimize the design  of the observational missions. Having taken all this into account, we have been led to consider the following two redshift ranges: $z\in [0,2.5]$ covering the available $H(z)$ expansion rate data, and $z\in [0,5]$, an extended redshift range to cover future missions. 

We will  report all the results obtained, and not just the best ones, in order to properly stress the strongest and also the weakest aspects of our approach. One should note that our conclusions will be based on two models, only. But the procedure can be extended, revealing the full power of the approach. For the moment, the main message of our study indicates that General Relativity with standard matter fields and in the presence of a quintessence field, $\phi$, is most likely in tension with the recent form of the string Swampland criteria, Eqs.~(\ref{eq:SC1}) and~(\ref{eq:SC2}), and that this situation will not change significantly with higher redshift data. Moreover, we have got a resonable confirmation that our approach here can indeed serve as a tool for deciding if the model is in the Swampland or not.

The paper is organized as follows. In Sect.~\ref{sec:Mod} we introduce the models and obtain the equations that explicitly show how the  generative process based on each model is explicitly constructed. On the way, we briefly discuss the most crucial aspects of the Bayesian Machine Learning approach, showing why it is an attractive method and how it can be used to constrain cosmological models by involving string Swampland criteria. We should remark that all this is presented in a very compact form, addressing only some crucial aspect of the philosophy behind the method and referring the readers to available references for more details. In Sect.~\ref{sec:results} we discuss the results obtained for the two redshift ranges $z\in [0,2.5]$ and $z\in [0,5]$, the extended one. The first covers recently available $H(z)$ data for the expansion rate, useful to validate the results of our fit. The other redshift range is designed to cover the results obtained in  future observational projects. Finally, in Sect.~\ref{sec:conc} we summarize the results of our analysis, followed by a discussion and final conclusions.

\section{The model and method}\label{sec:Mod}

In this section we will introduce the models and the method we have used in our analysis. Our starting point is General Relativity with standard matter fields in the presence of a quintessence field, $\phi$, considered  to be as the EFT, in the philosophy explained in the previous section. It is known that, in this case, the EFT will be described by the following action
\begin{equation}\label{eq_MT}
S = \int d^{4} x \sqrt{-g} \left( \frac{M^{2}_{P}}{2} R - \frac{1}{2} \partial_{\mu} \phi \partial^{\mu} \phi -V(\phi) \right ) + S_{m},
\end{equation}
where $S_{m}$ corresponds to standard matter, $M_{P} = 1/\sqrt{8\pi G}$ is the reduced Planck mass, $R$  the Ricci scalar, $\phi$ the field, and $V(\phi)$ the field potential. On the other hand, when we consider the FLRW universe, the dynamics of the scalar field's dark energy and dark matter will be described by the equations 
\begin{equation}\label{eq:drhoPi}
\dot{\rho}_{\phi} + 3 H (\rho_{\phi} + P_{\phi}) = 0,
\end{equation}
\begin{equation}\label{eq:drhoDm}
\dot{\rho}_{dm} + 3 H \rho_{dm} = 0.
\end{equation}
Moreover, Eqs.~(\ref{eq:drhoPi}) and~(\ref{eq:drhoDm}) are the energy conservation laws for the components describing the background dynamics\footnote{The form of   Eqs.~(\ref{eq:drhoPi}) and~(\ref{eq:drhoDm}) shows the absence of a coupling between the scalar field's dark energy and dark matter. In general,  a  coupling modifying the right hand side of Eqs.~(\ref{eq:drhoPi}) and~(\ref{eq:drhoDm}) is present, which corresponds to an interaction between dark energy and dark matter}. Furthermore, $\rho_{\phi}$, $\rho_{dm}$ and $P = P_{\phi}$ are related to each other through the Friedmann equations, as follows
\begin{equation}\label{eq:F1}
H^{2} = \frac{1}{3} (\rho_{\phi} + \rho_{dm}), 
\end{equation}
and 
\begin{equation}\label{eq:F2}
\dot{H} + H^{2} = -\frac{1}{6} (\rho_{\phi} + \rho_{dm} + 3 P_{\phi}).
\end{equation}

Now, when we assume that the scalar field is spatially homogeneous, for its energy density and pressure we have
\begin{equation}\label{eq:rhoPhi}
\rho_{\phi} = \frac{1}{2} \dot{\phi} + V(\phi),
\end{equation}
and
\begin{equation}\label{eq:pPhi}
P_{\phi} = \frac{1}{2} \dot{\phi} - V(\phi),
\end{equation}
where the dot means derivative w.r.t to the cosmic time, while $V(\phi)$ is the potential of the scalar field. In all equations  above, $H = \dot{a}/a$ is the Hubble parameter. On the other hand, it is easy to see, from Eqs.~(\ref{eq:rhoPhi}) and~(\ref{eq:pPhi}), that
\begin{equation}\label{eq:dphi2}
\dot{\phi}^{2} = \rho_{\phi} + P_{\phi},
\end{equation} 
while
\begin{equation}\label{eq:Vphi}
V(\phi) = \frac{\rho_{\phi} - P_{\phi}}{2}.
\end{equation}    
Now, as from Eq.~(\ref{eq:drhoDm}) we have $\rho_{dm} = 3 H_{0}^{2} \Omega_{0} (1+z)^{3}$, then from Eq.~(\ref{eq:F1}) we can determine the energy density of the scalar field, which in this case reads 
\begin{equation}
\rho_{\phi} = 3 H^{2} - 3 H_{0}^{2} \Omega_{0} (1+z)^{3},
\end{equation}
where $H_{0}$ is the Hubble parameter value at $z=0$~($z$ being the redshift). After some algebra, we  get that $P_{\phi} = 2(1+z)H H^{\prime} - 3 H^{2}$, where the prime denotes  derivative wrt the redshift. On the other hand, we  see immediately that $\rho^{\prime}_{\phi} = 6 H H^{\prime} - 9  H_{0}^{2} \Omega_{0} (1+z)^{2}$ and $P^{\prime}_{\phi} = 2 (1+z) (H^{\prime 2} + H H^{\prime \prime}) - 4 H H^{\prime}$. Coming back to the form of $SC1$ and $SC2$ to be used in the model based generative process, we need only take into account that $d V(\phi)/d\phi = (dV/dz)/(d\phi/dz)$, where $d\phi/dz$ is calculated from Eq.~(\ref{eq:dphi2}), and that $\dot{\phi} = -(1+z)H \phi^{\prime}$. This was the starting point of~\cite{SCGP}, where the authors used the reconstructed $H$, $H^{\prime}$ and $H^{\prime\prime}$ to do model-independent analysis and estimations. 

In this paper we follow to the same equations as  discussed above, but use instead new parameterizations for $H(z)$. In particular, we consider first a parameterization for the Hubble parameter, while in the second case we use a corresponding one for the deceleration parameter. Now, since we have at hand already the last element required to organize the generative process based on the model, let us pause a bit, to understand what Bayesian Machine Learning is, and how it may allow one to constrain cosmological models directly from the string Swampland criteria. To be mentioned, to start, is that the Bayesian Machine Learning approach is a likelihood-free inference method, which allows to perform Bayesian Inference with no reference to a likelihood function, and it is solely based on the forward simulations. Moreover, in likelihood-free inference  parameters are drawn from the prior and forward simulated mock data. In other words, in our analysis 

\begin{itemize}

\item We define the model, to be used to provide a so-called generative process. In our case it will be the string Swampland criteria.

\item We envisage the data to be the data obtained from the generative process\footnote{We address the readers to \cite{PyMC3_p1} and \cite{PyMC3_p2} for more details about the difference between Machine Learning and Bayesian Machine Learning that rose namely in this step}. 

\item We run the learning algorithm, to get a brand new distribution over the model parameters and update our prior believe. 

\end{itemize}

It should be mentioned that, within the adopted approach, one can break some of the usual limits in science. On the other hand, it gives a nice opportunity to perform more complex studies in cosmology and astrophysics, among other relevant research fields. We already mentioned above, that we just give here a very short description of the philosophy and do not go into any other details related with the way how one can device a model-based generative process. This process itself is already of much interest for programmers and for computer science and data science researchers; but for us it is more important to concentrate all our attention on the physics behind the problem. In this regard, we should mention that our search shows that there are several perfectly organized frameworks dedicated to this task and a particular one is very valuable for our case, when we want to concentrate attention on the physics only. After some tests, we eventually  found that it is best to use PyMC3 probabilistic programming, a python-based framework, which in practice has proven to be very fast, useful, and able to be easily integrated with other python-based frameworks~\cite{PyMC3}. It uses Theano\footnote{http://deeplearning.net/software/theano}, that is, a deep learning python-based library, to construct probability distributions and implement cutting edge inference algorithms. We do have an additional reason to have omitted any specific discussion on the mathematical framework behind deep learning algorithms and Bayesian Machine Learning: the PyMC3 standard tutorials are endowed with numerous excellent examples demonstrating how the above ideas can actually be implemented within a practical problem.

\section{Numerical results and analysis}\label{sec:results}

In this section we will discuss the results. For convenience, we have divided into two subsections. Before starting with the presentation and discussion of our results we should mention that our initial study shows that both string Swampland criteria, namely $SC1$, Eq.~(\ref{eq:SC1}), and $SC2$, Eq.~(\ref{eq:SC2}), can be used in our generative process. Moreover, in both cases we get well consistent results. However, it is reasonable to concentrate on the generative process based on $CS1$, Eq.~(\ref{eq:SC1}), the reason being that the generative process is computationally less costly in this case than if we use $SC2$, Eq.~(\ref{eq:SC2}). This can be also seen from the structure of the problem and the number of mathematical expressions to be evaluated in order to use $SC2$.

\subsection{Model with a given $H(z)$ parameterization}

The first model constrained in this work, using the string Swampland criterion $SC1$, Eq.~(\ref{eq:SC1}), is described by the following expansion rate 
\begin{equation}\label{eq:H1}
H(z) = H_{0} + H_{1}\frac{z^{2}}{1+z},
\end{equation}
where $H_{0}$ and $H_{1}$ are two parameters to be determined, while $z$ is the redshift. In other words this is a model independent parameterization of the Hubble parameter and the form has been found from more general parameterizations using the Bayesian Machine Learning approach. Once the form of $H(z)$ is known, we can calculate higher-order derivatives and use them to start the analysis of the model by employing the $SC1$ criterion (owing to the reasons mentioned at the beginning of this section). 

\begin{figure}[h!]
 \begin{center}$
 \begin{array}{cccc}
\includegraphics[width=90 mm]{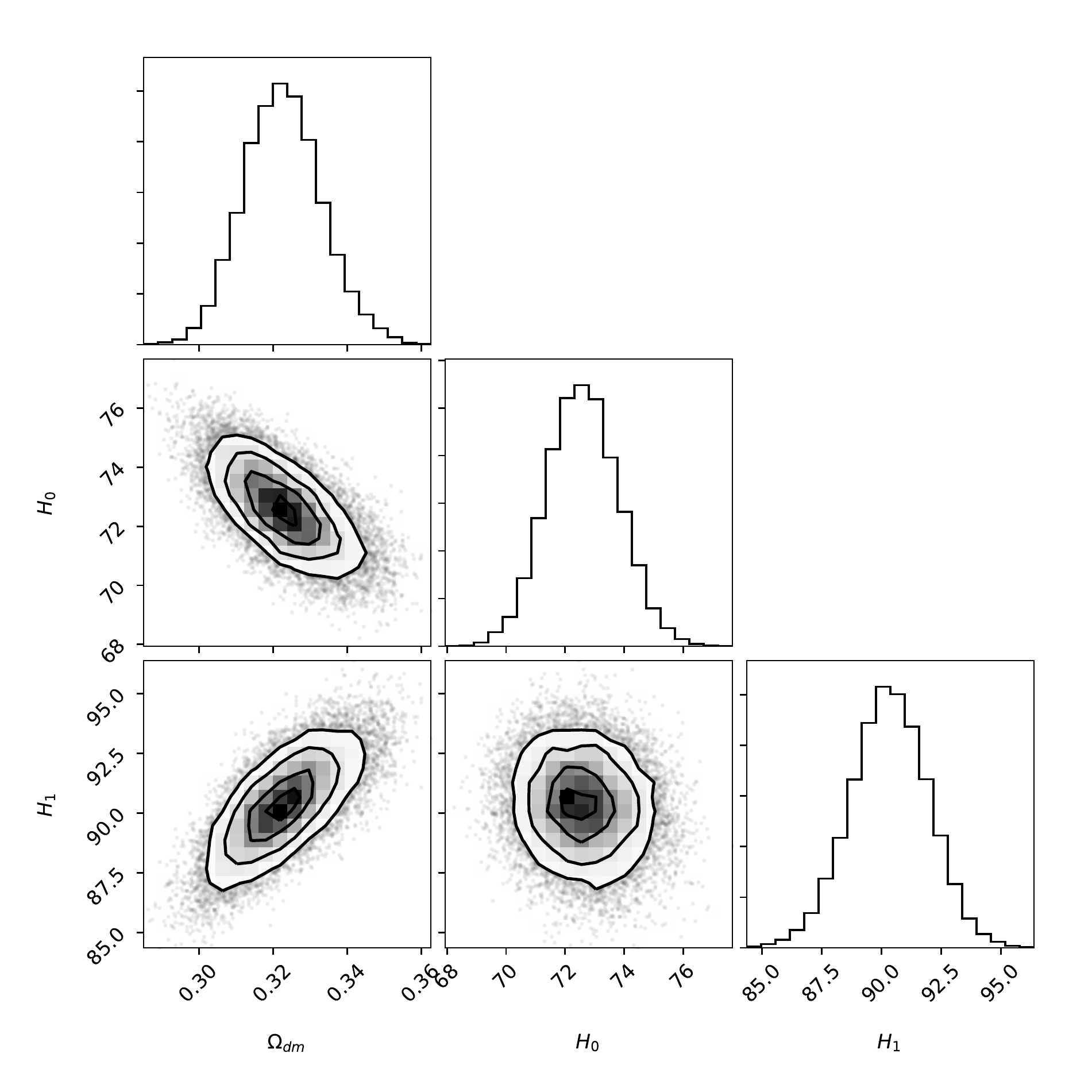}
\includegraphics[width=90 mm]{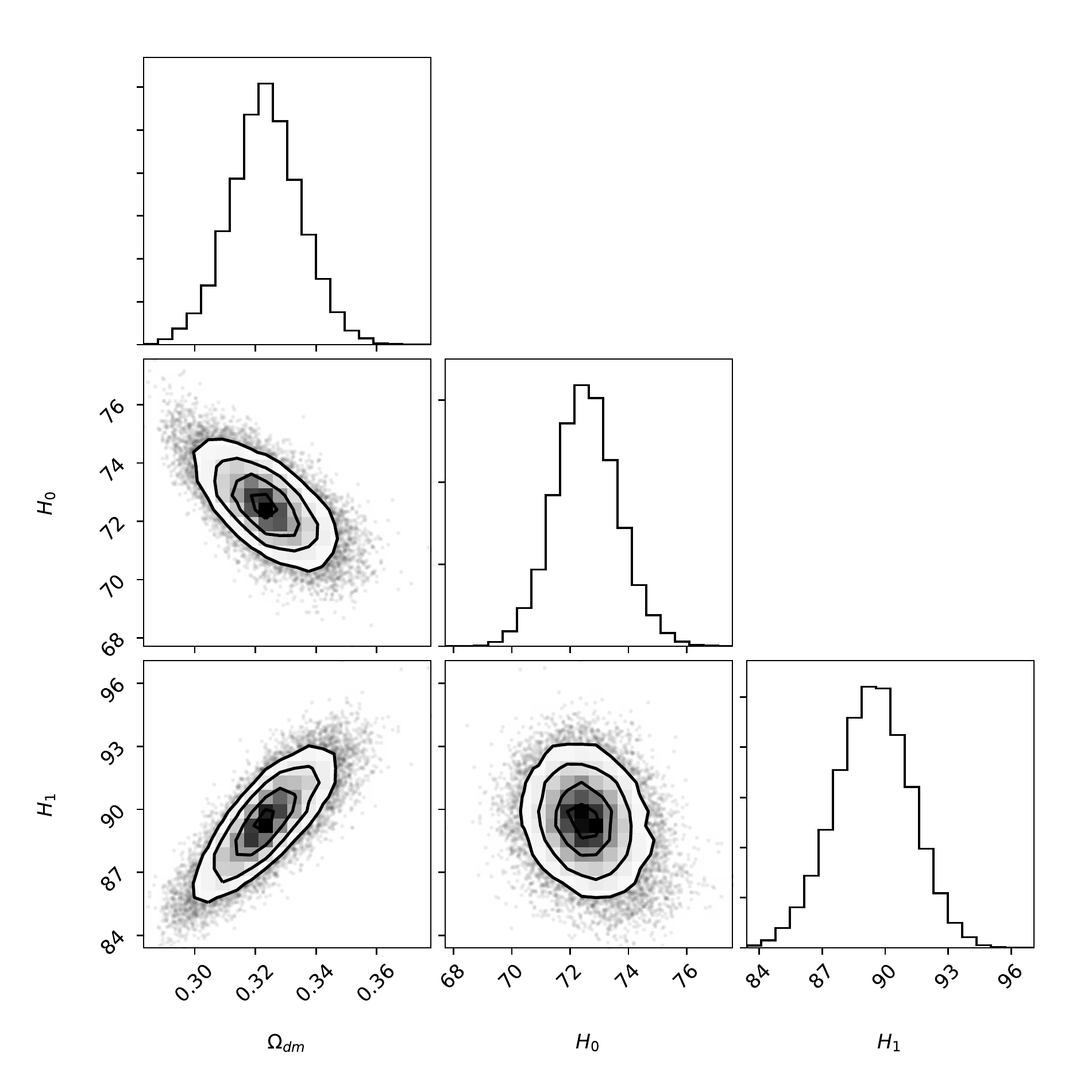}
 \end{array}$
 \end{center}
\caption{The contour map of the model given by Eq.~(\ref{eq:H1}), for $z \in [0,2.5]$, is given on the left hand side plot. The one on the right hand side depicts the contour map of the same model for $z \in [0,5]$. The best fit values of the model parameters are $\Omega_{dm} = 0.322 \pm 0.01$, $H_{0} = 72.59 \pm 1.21$  km/s/Mpc and $H_{1} =  90.34 \pm 1.61$ km/s/Mpc for $z\in[0,2.5]$. For $z \in [0,5]$, they are found to be $\Omega_{dm} = 0.325 \pm 0.012$, $H_{0} = 72.55 \pm 1.06$ km/s/Mpc and $H_{1} = 89.43 \pm 1.8$ km/s/Mpc. The generative process has been constructed using SC1, Eq.~(\ref{eq:SC1}), where $H(z)$ given by Eq.~(\ref{eq:H1}) has been used to get $H^{\prime}$ and $H^{\prime \prime}$.}
 \label{fig:Fig1_0}
\end{figure}

The contour maps of the model given by Eq.~(\ref{eq:H1}), for $z \in [0,2.5]$ and $z \in [0,5]$, respectively, can be found in Fig.~(\ref{fig:Fig1_0}). We observe that:

\begin{itemize}

\item When we consider $z \in [0,2.5]$, then the best fit values and $1\sigma$ errors are $\Omega_{dm} = 0.322 \pm 0.01$, $H_{0} = 72.59 \pm 1.21$  km/s/Mpc and $H_{1} =  90.34 \pm 1.61$ km/s/Mpc.  

\item When we consider $z \in [0,5]$, then the best fit values and $1\sigma$ errors are $\Omega_{dm} = 0.325 \pm 0.012$, $H_{0} = 72.55 \pm 1.06$ km/s/Mpc and $H_{1} = 89.43 \pm 1.8$ km/s/Mpc.

\end{itemize}

From these results, it is clear that Bayesian Machine Learning using $SC1$, Eq.~(\ref{eq:SC1}), certainly imposes very tight constraints on $\Omega_{dm}$ at $z=0$. Moreover, we observe that, most likely, higher redshift related field traverse will not affect the constraints on $\Omega_{dm}$ at $z=0$. Another interesting result that immediately captures our attention is the constraint obtained for $H_{0}$, which according to the parametrization considered, corresponds to the Hubble parameter at $z=0$. In particular, we found that the $H_{0}$ tension can be solved within $1 \sigma$. In other words, the model can be useful to alleviate the $H_{0}$ tension problem efficiently\footnote{We refer to \cite{PyMC3_p1} and \cite{PyMC3_p2} for more details about the $H_{0}$ tension problem and about the approaches that may lead to its solution}.  We see also that, most likely, higher redshift related field traverse will affect the constraints on $H_{0}$ and $H_{1}$.

\begin{figure}[t!]
 \begin{center}$
 \begin{array}{cccc}
\includegraphics[width=80 mm]{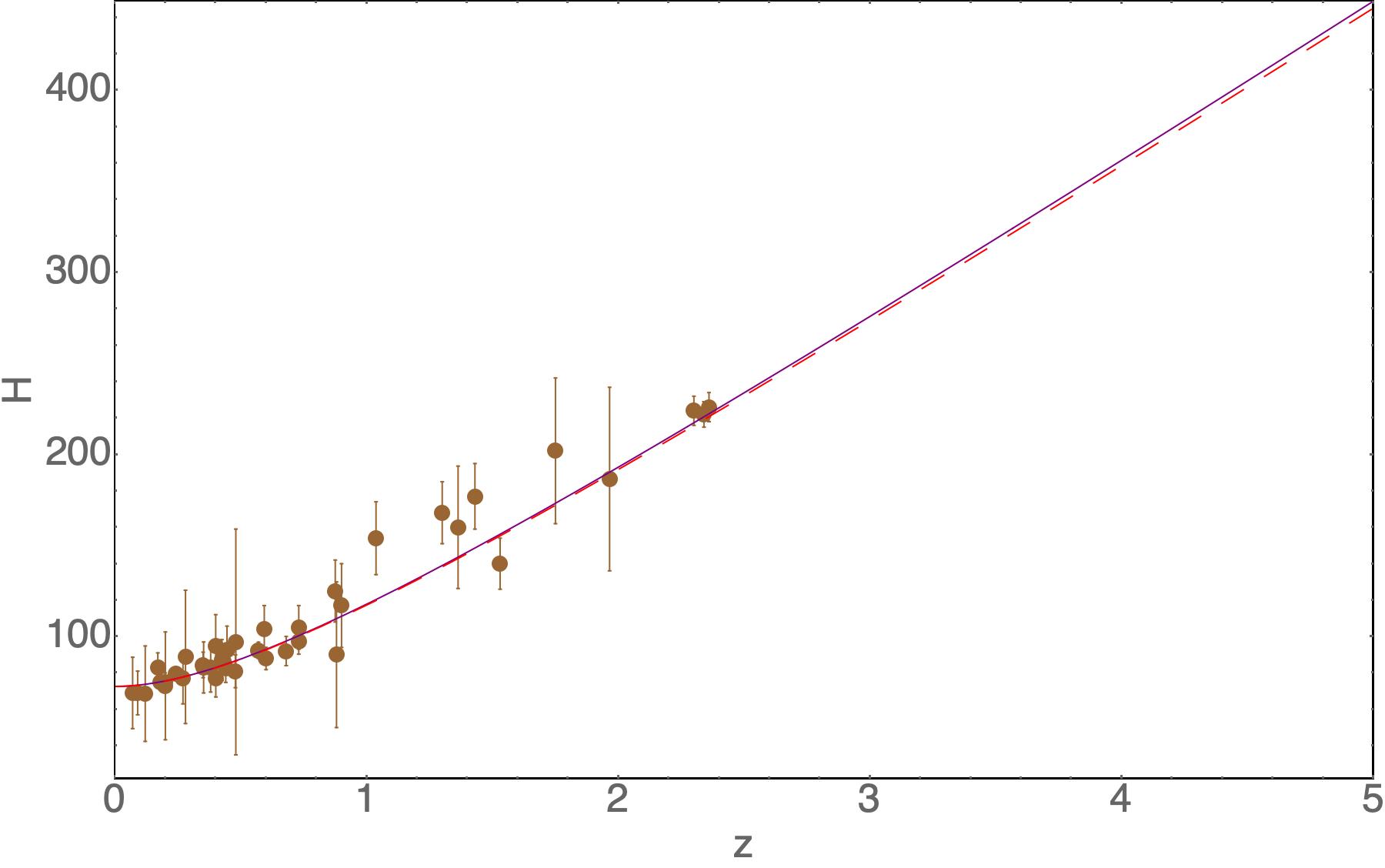}&&
\includegraphics[width=80 mm]{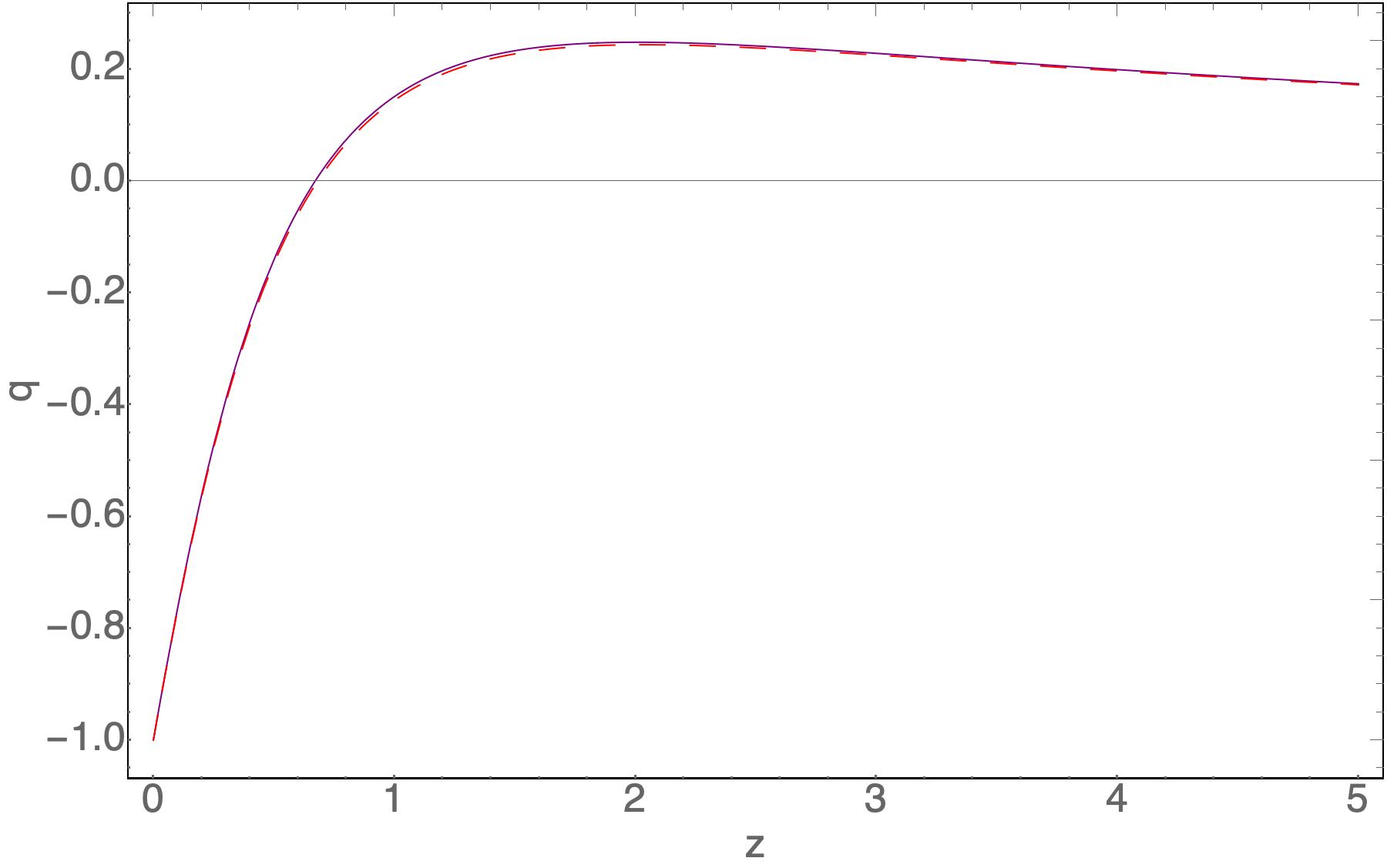}\\
\includegraphics[width=80 mm]{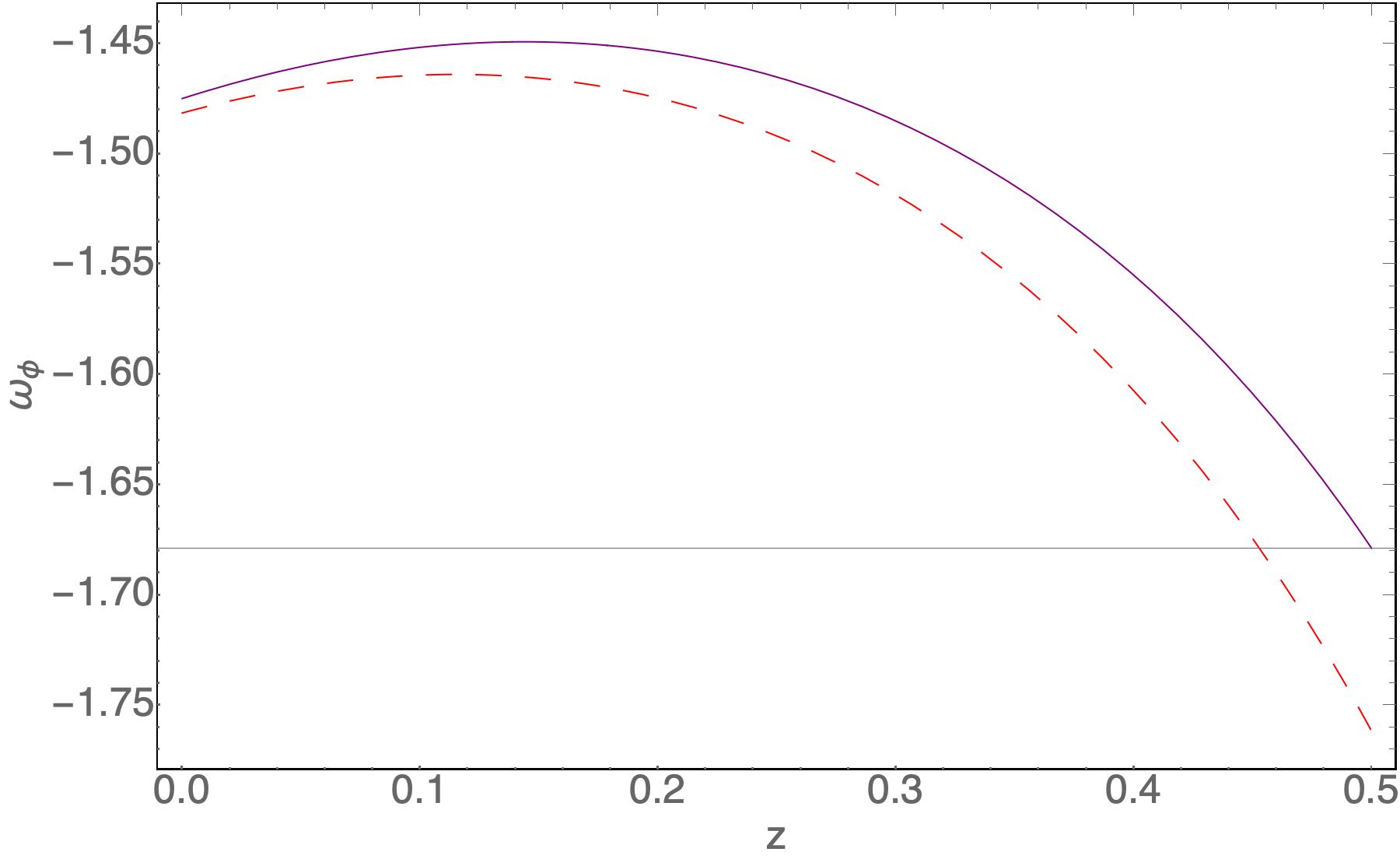}&&
\includegraphics[width=80 mm]{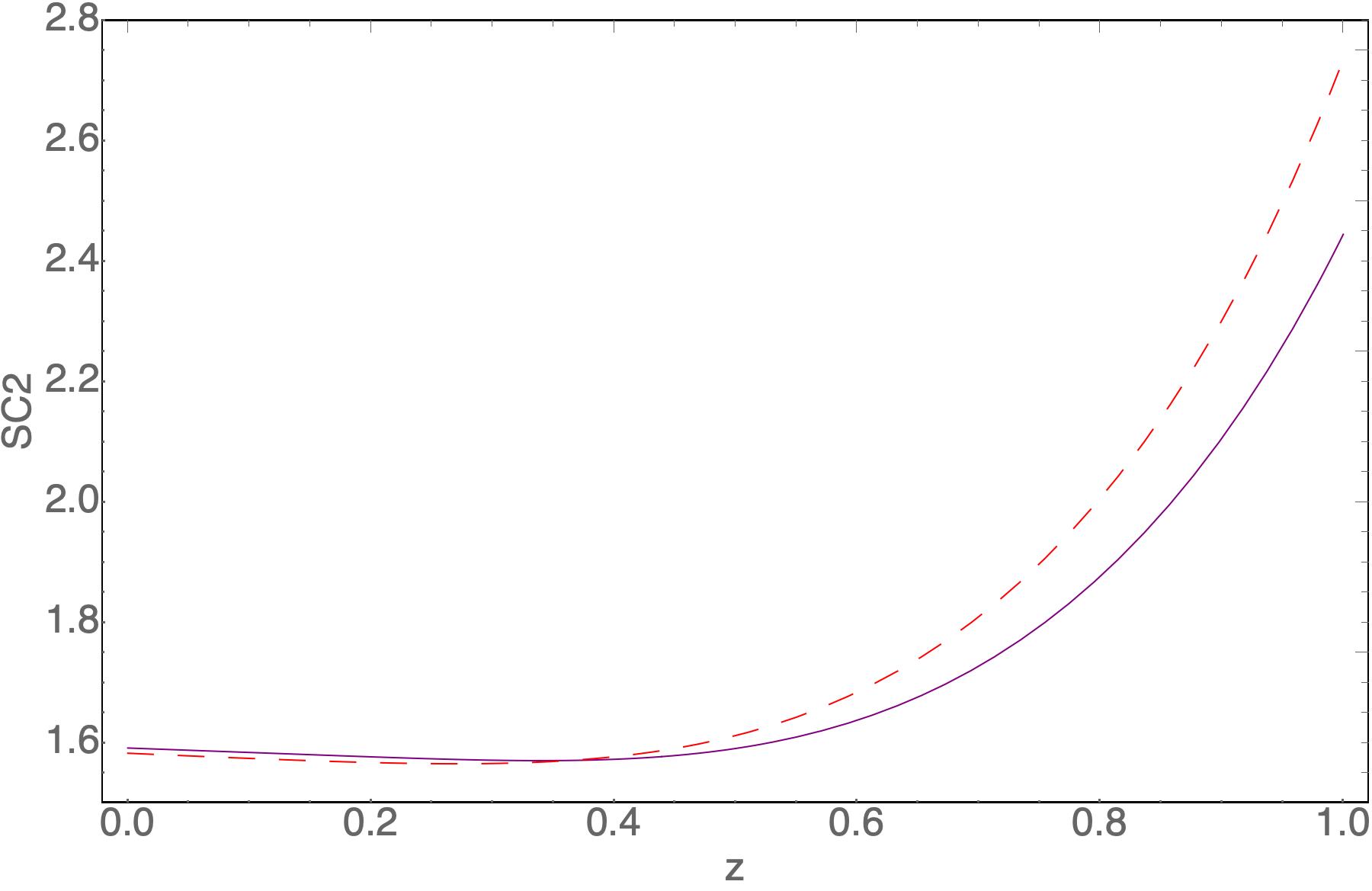}
 \end{array}$
 \end{center}
\caption{The graphical behavior of the Hubble parameter in comparison with known $H(z)$ data is presented on the left hand side plot of the top panel. The plot on the right hand side corresponds to the behavior of the deceleration parameter. The left hand side plot of the bottom panel depicts the behavior of the $\omega_{\phi} = P_{\phi}/\rho_{\phi}$ equation of state parameter describing the scalar field dark energy model. The right hand side plot of the bottom panel shows the graphical behavior of SC2 Eq.~(\ref{eq:SC2}). In all cases the purple curve is a plot for the best fit values of the model parameters, when $z \in [0,2.5]$, while the dashed red curve has been chosen for the case when $z \in [0,5]$. The red dots represent the known observational $H(z)$ data,  the same as in Ref.~\cite{SCGP}. The redshift range for $\omega_{\phi}$ has been chosen in order to make it possible to compare our results with those presented in~\cite{SCGP}. The Bayesian Machine Learning procedure has been devised for the model given by Eq.~(\ref{eq:H1}). The best fit values used in the analysis can be found in Table~\ref{tab:Table1}.}
 \label{fig:Fig1_1}
\end{figure}

From our analysis, we can extract some additional messages. First, we see that the model considered does in fact explain the accelerated expansion of our Universe, and clearly indicates the transition to this evolution phase, as well. Moreover, using the best fit values of the parameters of the model, the top-right plot of Fig.~(\ref{fig:Fig1_1}) shows that the model can also explain the BOSS result for the Hubble parameter, at $z=2.34$. 

Now, before presenting the second part of our observations, we would like to highlight once more that, starting from $SC1$, Eq.~(\ref{eq:SC1}), we have constrained the background dynamics of the model described by Eq.~(\ref{eq_MT}) and replaced the scalar field by a convenient parameterization of the Hubble constant. This allows also to estimate the behavior of the $\omega_{\phi} = P_{\phi}/\rho_{\phi}$ and $SC2$, Eq.~(\ref{eq:SC2}). The results can be found on the bottom panel of Fig.~(\ref{fig:Fig1_1}) indicating that, at low redshift, a dark energy dominated Universe will not end up in the Swampland. However, the behavior of $\omega_{\phi} = P_{\phi}/\rho_{\phi}$ gives hints on the tension that may arise between string Swampland criteria and General Relativity with standard matter and quintessence dark energy. Since we found that, according to the best fit values of the model parameters, dark energy has a phantom nature. However, it should be also mentioned, that  another interesting massage arises from our study, this time indicating that it is possible to avoid being in the Swampland; although in this case dark energy should have a phantom nature. In a way, this should be not surprising at all. Moreover, a hint about a possible tension between Swampland criteria and $H(z)$ data can be guessed from the second part of Table~\ref{tab:Table1}, where the constraints on the $H_{0}$ and $H_{1}$ parameters have been obtained using a generative process based on Eq.~(\ref{eq:H1}).  

To end this subsection, let us summarize wherefrom we started and what we have accomplished at this point. Specifically, first of all, we have proven that, incorporating both Bayesian Machine Learning and string Swampland criteria, we are definitely able to constrain cosmological models. Then, we have validated the fit results using observational expansion rate data already available. Moreover, our procedure allowed to demonstrate that the string Swampland criteria is in a tension with recent observations and the understanding of dark energy, in the scheme of General Relativity with standard matter and a dark energy splitting approach. Moreover, we noticed a spontaneous sign switch in the dark energy equation of state parameter, when the scalar field net excursion leaves the domain $z\in [0,2.5]$.
    
On the other hand, our results come from a specific form of the expansion rate, constructed in the Bayesian Machine Learning approach; therefore, additional study is required to see if this conclusion depends on the procedure. However, it must be stressed once more that, in the  generative based Bayesian Machine Learning procedure employed, we did not use any observational data. In this regard, it would be interesting to investigate other expansion rate models, to reveal if and how the specific Bayesian Machine Learning approach being used will affect the Swampland related results. Aiming  at this goal, we consider in addition another toy model, in this case however, a model for the deceleration parameter parameterization. Results for this case are described in the next subsection.

\begin{table}
  \centering
    \begin{tabular}{ | c | c | c | c |  p{2cm} |}
    \hline
    
  $H(z) = H_{0} + H_{1} \frac{z^{2}}{1+z}$  & $\Omega_{dm}$ & $H_{0}$ & $H_{1}$  \\
      \hline
      
 when $z\in[0,2.5]$ & $0.322 \pm 0.01$ & $72.59 \pm 1.21$ km/s/Mpc & $90.34 \pm 1.61$ km/s/Mpc\\
          \hline
          
when $z\in[0,5]$  & $0.325 \pm 0.012$ & $72.55 \pm 1.06$ km/s/Mpc & $89.43 \pm 1.8$ km/s/Mpc \\
         \hline
        
 \multicolumn{3}{c}{} \\ \hline
 
$H(z) = H_{0} + H_{1} \frac{z^{2}}{1+z}$ & $\Omega_{dm}$ & $H_{0}$ & $H_{1}$ \\
       \hline

when $z\in[0,2.5]$ & $-$ & $71.524 \pm 0.5$ km/s/Mpc & $90.654 \pm 0.5 $ km/s/Mpc  \\
         \hline
 when $z\in[0,5]$  & $-$ & $71.094 \pm 0.5$ km/s/Mpc & $91.83 \pm 0.1$ km/s/Mpc  \\  
 
     \hline 
    
    \end{tabular}
\caption{Best fit values and $1\sigma$ errors estimated for Model 1, Eq.~(\ref{eq:H1}), for $z \in [0,2.5]$ and for $z \in [0,5]$, respectively. The results depicted in the first part of the table are obtained from the generative process based on SC1, Eq.~(\ref{eq:SC1}). The second part of the table corresponds to the results of the fit when the theoretical form of the expansion rate, Eq.~(\ref{eq:H1}), is used in the generative process.}
  \label{tab:Table1}
\end{table}

\subsection{Model with a given $q(z)$ parameterization}

The second toy model we are going to consider is based on a deceleration parameter parameterization, given by the following expression~\cite{qz}
 \begin{equation}\label{eq:q1}
 q(z) = \frac{1}{2} +  \frac{q_{1} z + q_{2}}{(z+1)^2},
\end{equation}
where $q_{1}$ and $q_{2}$ are the free parameters to be determined. It is easy to see, that the expansion rate of the Universe with such a deceleration parameter has the  form
\begin{equation}\label{eq:q1_H}
H(z) = H_{0} (z+1)^{3/2} e^{\frac{z (q_{1} z + q_{2} (z+2))}{2 (z+1)^2}},
\end{equation}
where $H_{0}$, $q_{1}$ and $q_{2}$ are the parameters of the model, this one having one parameter more, as compared to the previous model. Following the same procedure as in case of the first model, we have been able to constrain the new parameters. In particular, we have found that

\begin{itemize}

\item When we consider $z \in [0,2.5]$, then the best fit values and $1\sigma$ errors are $\Omega_{dm} = 0.322 \pm 0.01$, $H_{0} = 73.64 \pm 1.5$ km/s/Mpc, $q_{1} =  -0.27^{+0.181}_{-0.172}$  $q_{1} = -1.48 \pm 0.03$.

\item When we consider $z \in [0,5]$, then the best fit values and $1\sigma$ errors are $\Omega_{dm} = 0.298\pm 0.0167$, $H_{0} = 73.63^{+1.476}_{-1.52}$ km/s/Mpc and $q_{1} = -0.139^{+0.13}_{-0.127}$ and $q_{2} = -1.47 \pm 0.029$.

\end{itemize}

\begin{table}
  \centering
    \begin{tabular}{ | c | c | c | c |  p{2cm} |}
    \hline
    
$q(z) = \frac{1}{2} +  \frac{q_{1} z + q_{2}}{(z+1)^2}$ & $\Omega_{dm}$ & $H_{0}$ & $q_{1}$ & $q_{2}$ \\
       \hline

when $z\in[0,2.5]$ & $0.292\pm 0.0181$ & $73.64 \pm 1.5$ km/s/Mpc & $-0.27^{+0.181}_{-0.172} $  & $-1.48 \pm 0.03 $\\
         \hline
 when $z\in[0,5]$  & $0.298\pm 0.0167$ & $73.63^{+1.476}_{-1.52}$ km/s/Mpc & $-0.139^{+0.13}_{-0.127} $  & $-1.47 \pm 0.029 $\\  
 
     \hline 
     
     \multicolumn{3}{c}{} \\ \hline

$q(z) = \frac{1}{2} +  \frac{q_{1} z + q_{2}}{(z+1)^2}$ & $\Omega_{dm}$ & $H_{0}$ & $q_{1}$ & $q_{2}$ \\
       \hline

when $z\in[0,2.5]$ & $-$ & $73.23 \pm 0.23$ km/s/Mpc & $0.043 \pm 0.056$  & $-1.48 \pm 0.026 $\\
         \hline
 when $z\in[0,5]$  & $-$ & $73.56 \pm 0.05$ km/s/Mpc & $-0.141 \pm 0.0078$  & $-1.49 \pm 0.0054 $\\  
 
     \hline

    \end{tabular}
\caption{Best fit values and $1\sigma$ errors estimated for Model 2, Eq.~(\ref{eq:q1_H}), for $z \in [0,2.5]$ and for $z \in [0,5]$, respectively. The results presented in the first part of the table are obtained from the generative process based on SC1, Eq.~(\ref{eq:SC1}). The second part of the table represents the fit results when the theoretical form of the expansion rate, Eq.~(\ref{eq:q1_H}), is used in the generative process.}
  \label{tab:Table2}
\end{table}

\begin{figure}[h!]
 \begin{center}$
 \begin{array}{cccc}
\includegraphics[width=90 mm]{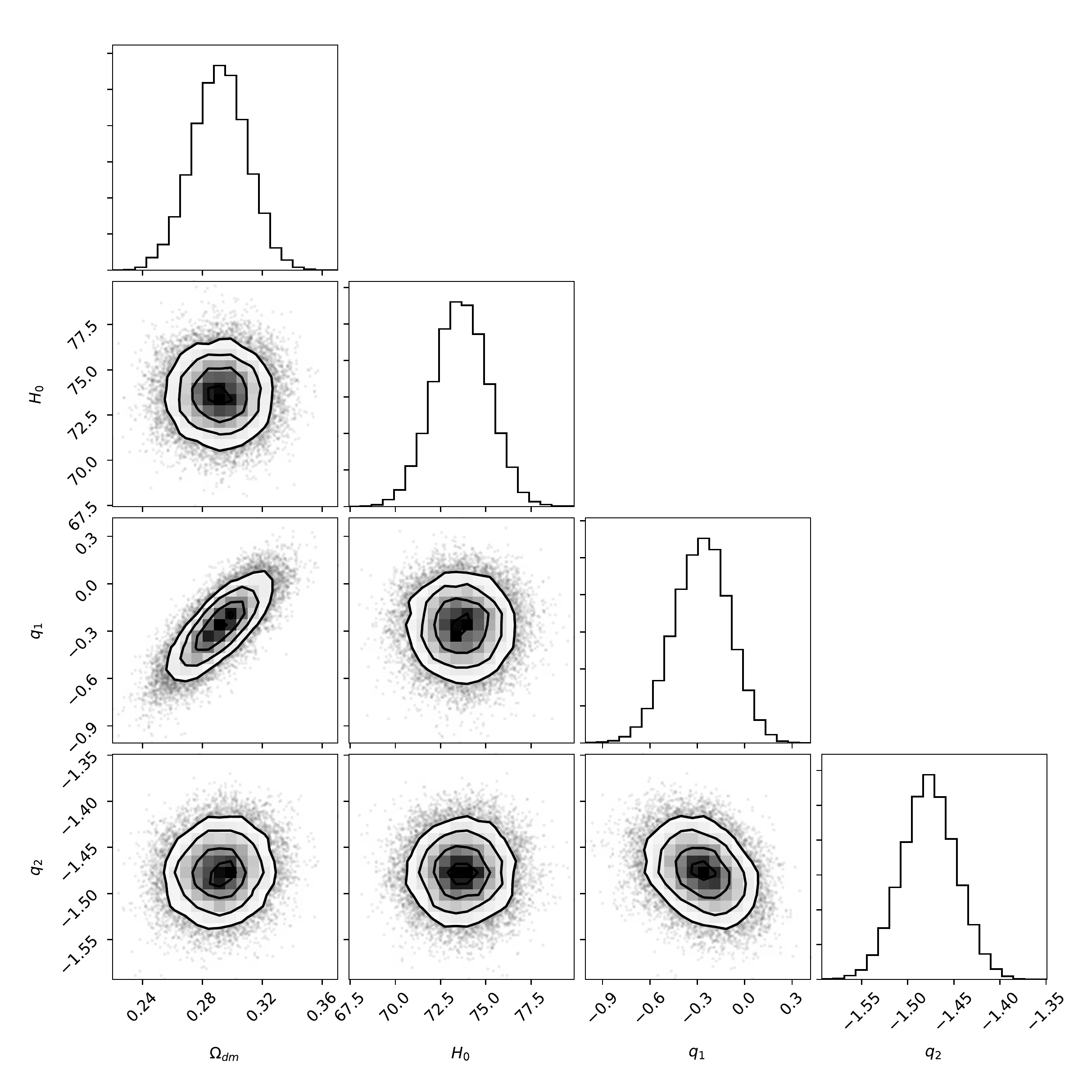}
\includegraphics[width=90 mm]{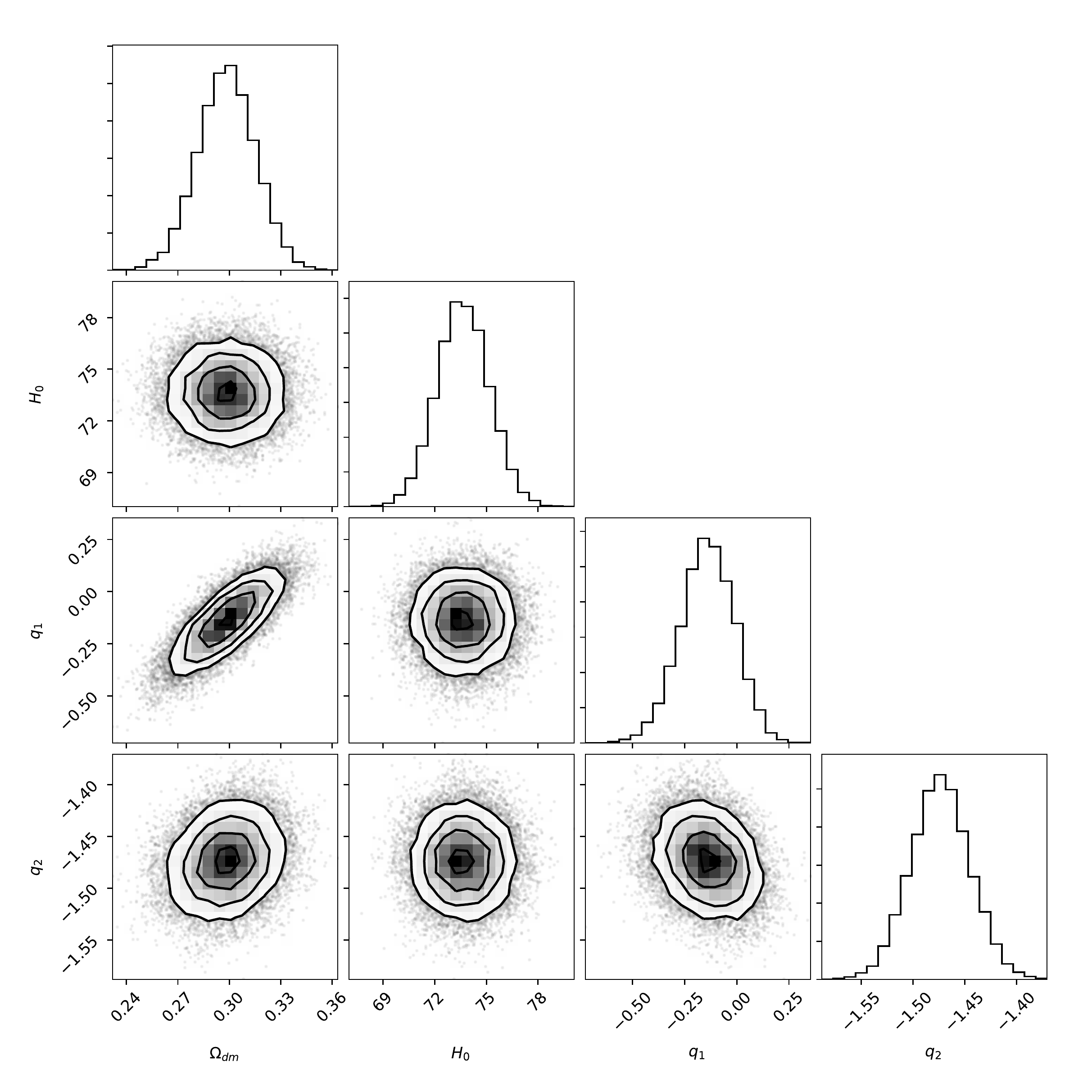}
 \end{array}$
 \end{center}
\caption{The contour map of the model given by Eq.~(\ref{eq:q1_H}), for $z \in [0,2.5]$, is given by the left hand side plot. The one on right hand side corresponds to the contour map of the same model, for $z \in [0,5]$. The best fit values of the model parameters are, respectively, $\Omega_{dm} = 0.322 \pm 0.01$, $H_{0} = 73.64 \pm 1.5$ km/s/Mpc, $q_{1} =  -0.27^{+0.181}_{-0.172}$,  $q_{1} = -1.48 \pm 0.03$, for $z\in[0,2.5]$, while for $z \in [0,5]$ they are found to be $\Omega_{dm} = 0.298\pm 0.0167$, $H_{0} = 73.63^{+1.476}_{-1.52}$ km/s/Mpc and $q_{1} = -0.139^{+0.13}_{-0.127}$ and $q_{2} = -1.47 \pm 0.029 $. The generative process has been constructed using SC1, Eq.~(\ref{eq:SC1}), where $H(z)$ is given by Eq.~(\ref{eq:q1_H}). It has been used to get $H^{\prime}$ and $H^{\prime \prime}$.}
 \label{fig:Fig2_0}
\end{figure}

\begin{figure}[t!]
 \begin{center}$
 \begin{array}{cccc}
\includegraphics[width=80 mm]{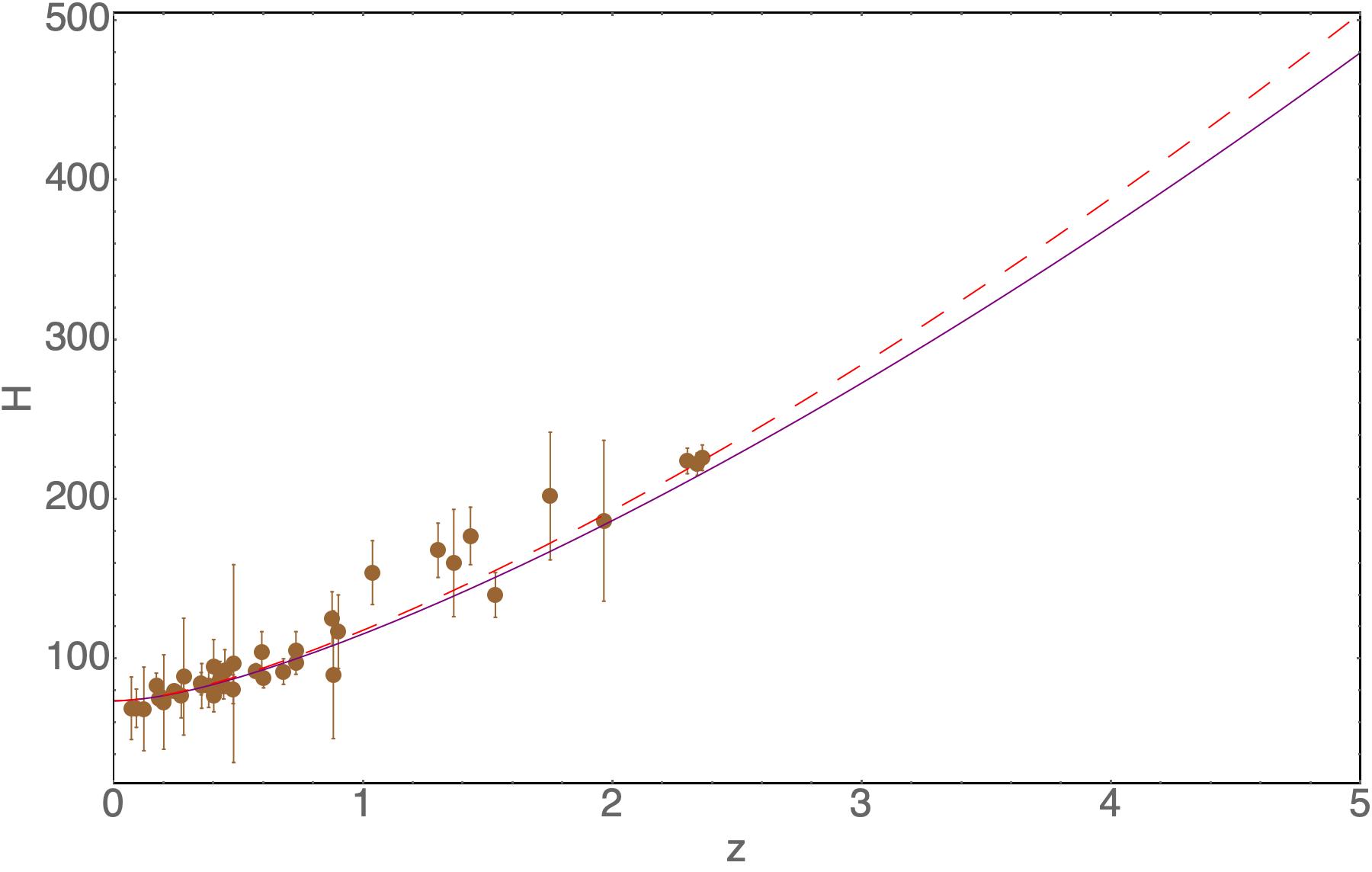}&&
\includegraphics[width=80 mm]{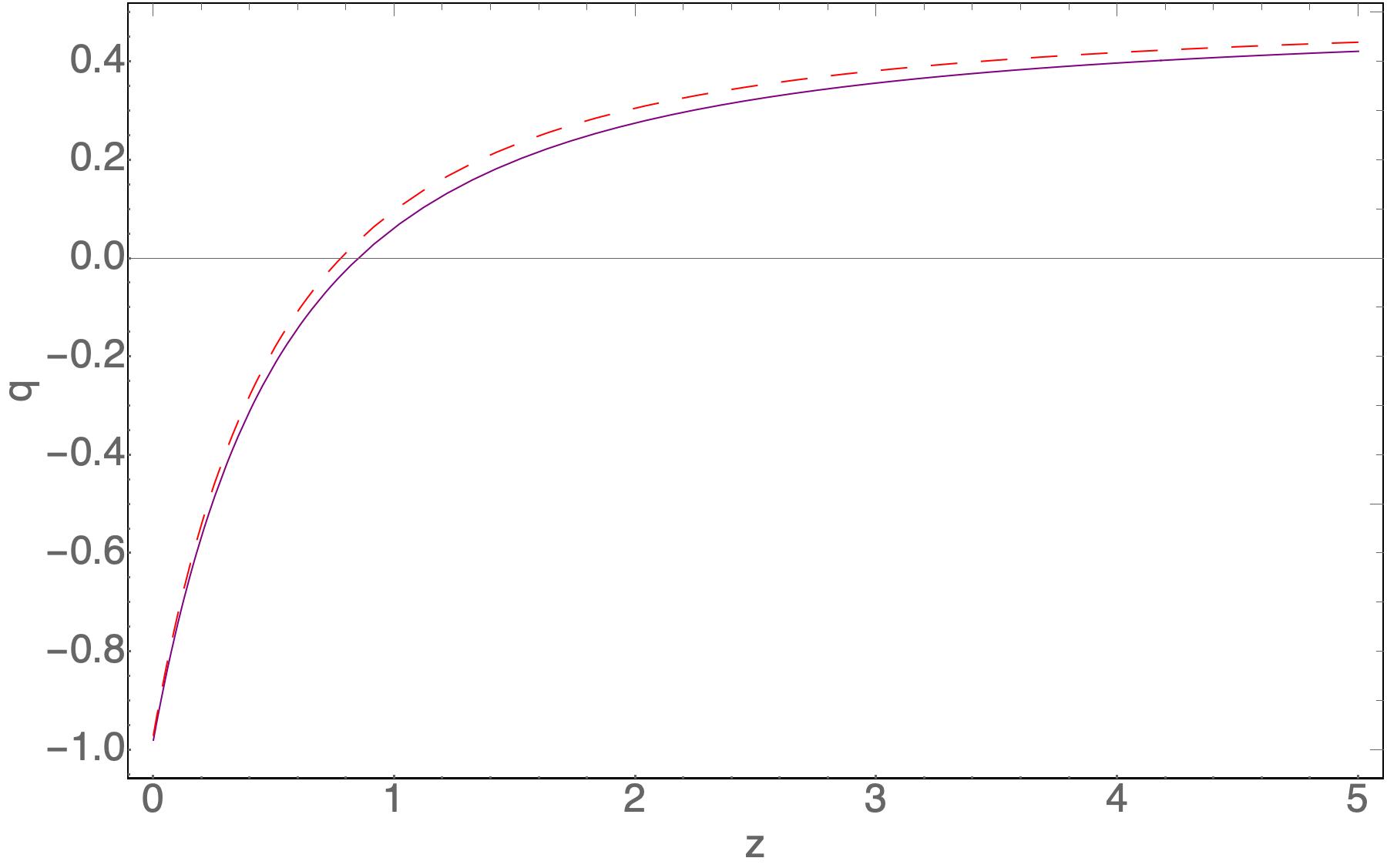}\\
\includegraphics[width=80 mm]{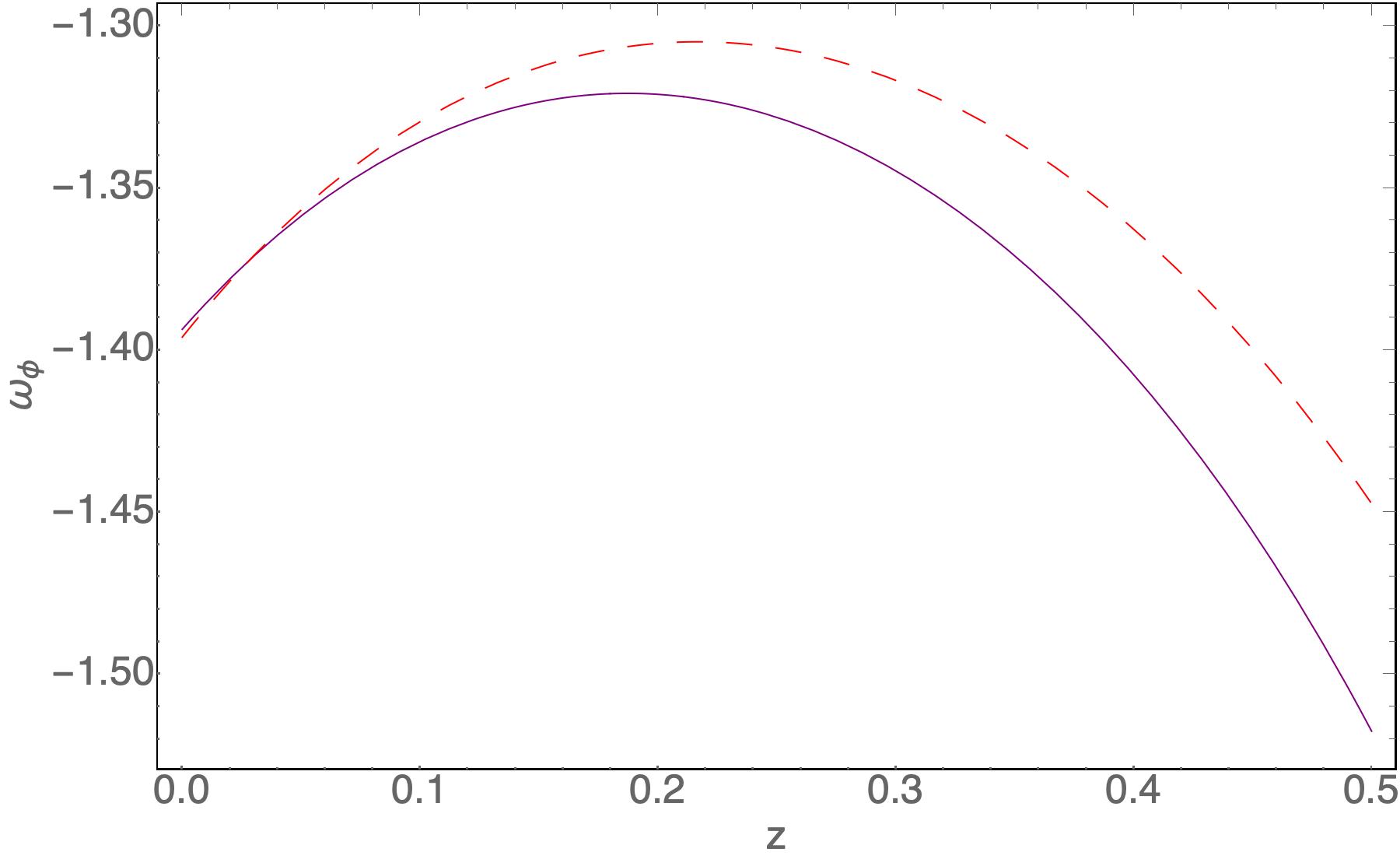}&&
\includegraphics[width=80 mm]{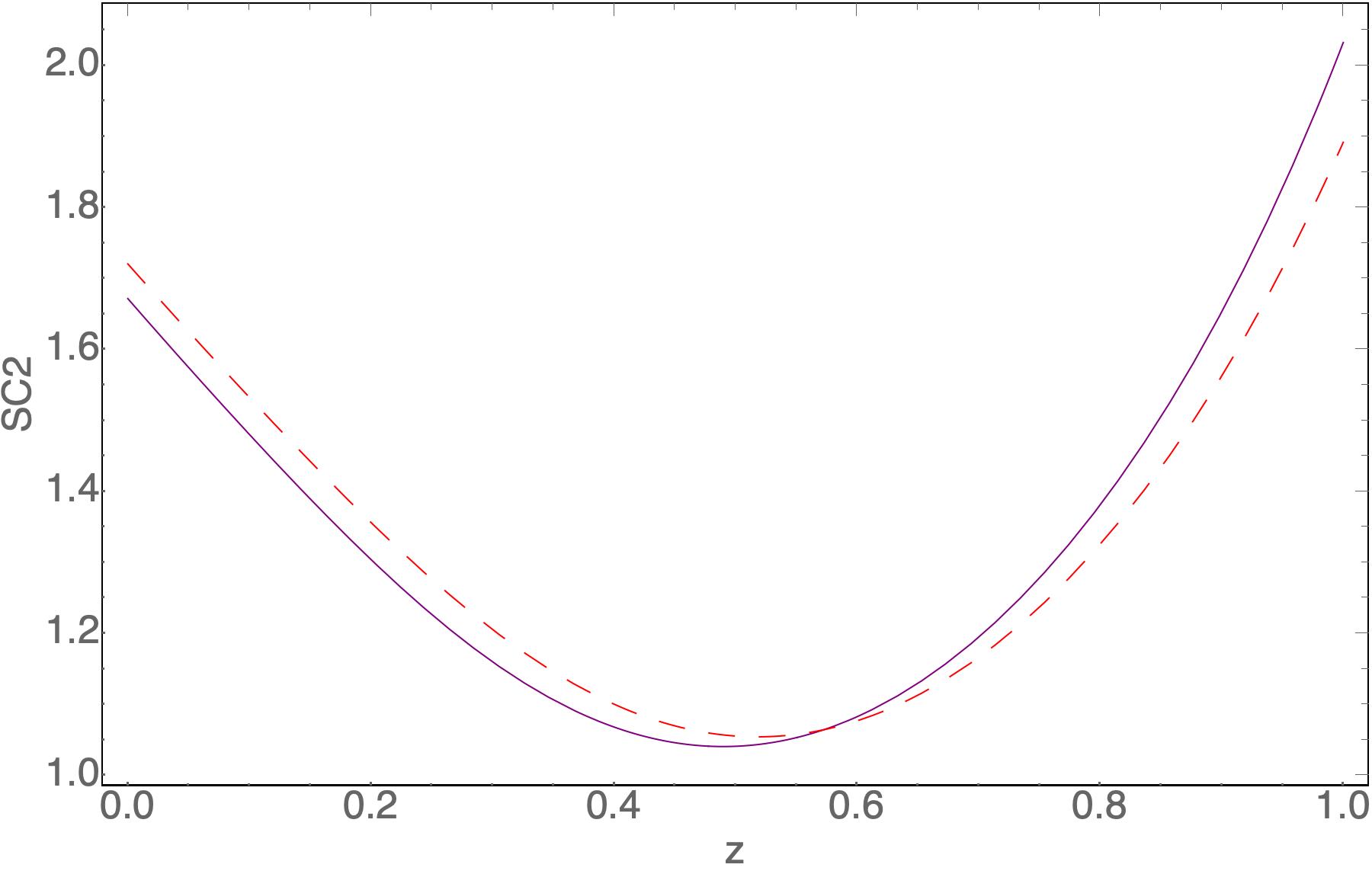}
 \end{array}$
 \end{center}
\caption{A graphical behavior of the Hubble parameter, compared with known $H(z)$ data, is depicted on the left hand side plot of the top panel. On the  right hand side plot, the behavior of the deceleration parameter is represented. The left hand side plot of the bottom panel corresponds to the behavior of the $\omega_{\phi} = P_{\phi}/\rho_{\phi}$ equation of state parameter describing the scalar field dark energy model. Further, the right hand side plot of the bottom panel shows the graphical behavior of SC2, Eq.~(\ref{eq:SC2}). In all cases, the purple curve is the plot for the best fit values of the  parameters of the model, when $z \in [0,2.5]$; while the dashed red curve corresponds to the case when $z \in [0,5]$. The red dots represent the known observational $H(z)$ data, and it is the same as in~\cite{SCGP}. The redshift range for $\omega_{\phi}$ has been chosen in order to make it possible to compare our results with those given in\cite{SCGP}. The Bayesian Machine Learning procedure has been constructed for the model given by Eq.~(\ref{eq:q1_H}). The best fit values used in the analysis are to be found in Table~\ref{tab:Table2}.}
 \label{fig:Fig2_1}
\end{figure}

The contour maps of the model given by Eq.~(\ref{eq:q1_H}), for $z \in [0,2.5]$ and $z \in [0,5]$, can be found in Fig.~(\ref{fig:Fig2_0}). Similar to the previous case, our results prove that the string Swampland criteria, Eq.~(\ref{eq:SC1}) and Eq.~(\ref{eq:SC2}), is in tension with General Relativity with standard dark matter and quintessence dark energy splitting. On the other hand, similarly to the previous case, we get the hint that it could be possible to build an effective theory not ending up in the Swampland, but then the dark energy in the low-redshift Universe should have phantom nature. Moreover, the behavior of the $\omega_{\phi} = P_{\phi}/\rho_{\phi}$ equation of state parameter, describing the scalar field dark energy model, would be in very good agreement with the results of~\cite{SCGP}, too. Moreover, during the validating of our fit results with the expansion rate data, we have found that, most likely, a high-redshift tension between the expansion rate, Eq.~(\ref{eq:q1_H}), based on theoretical results and observations, will arise. This has been observed when the field traverse has been limited to $z\in [0,2.5]$. On the other hand, when the field traverse extends to $z\in [0,5]$, the mentioned tension actually disappears. In other words, the model can explain the accelerated expansion of the low redshift Universe and the transition to that phase. Moreover, it is free from the $H_{0}$ tension problem due to the best fit values of the model parameters, and can explain the BOSS result for the Hubble parameter at $z=2.34$. 

All these  results can be seen in Fig.~(\ref{fig:Fig2_1}), where a comparison of the ones corresponding to $z\in [0,2.5]$, with those coming from $z\in [0,5]$, limiting the field traverse, can be found, too. In addition, the results summarizing how the  two redshift ranges considered for the field traverse will affect the model parameters constraints, can be found in the first part of Table~\ref{tab:Table2}. The second part of the table shows the corresponding constrains when the generative process is based on Eq,~(\ref{eq:q1_H}). This is another way of showing that the string Swampland criteria for a lower redshift dark energy dominated Universe is most likely in tension with the up to now available observations. 

To finish this subsection, we should stress again, that an interplay between Bayesian Machine Learning and the Swampland criteria can provide a working pipeline allowing to constrain the free parameters of the underlying theory. Moreover, it can identify the position of the theory in phase space, clearly showing if belongs (or not) to the Swampland  and, moreover, if it will, or not, end there. The two toy model discussed yield already clear enough results, which demonstrate how our approach can be used successfully, to extract valuable information using minimal information. Certainly, several other questions still arise, which should be clarified before a final conclusion can be reached, but it seems already clear that we have started to explore a very promising direction of study, which reveals a connection between Swampland and cosmology. Further results on this topic would be reported in forthcoming papers. Finally, similar to the previous model, we observe again a spontaneous sign switch in the dark energy equation of state parameter. This is an interesting phenomenon requiring further analysis, too.

\section{Conclusions}\label{sec:conc}

We say nothing new when stating that Machine Learning, and in general Artificial Intelligence, has brought new insights in data analysis and data engineering, computer, and vision science. Nowadays, thanks to the very robust algorithms developed, Machine Learning allows our computers to interact with us, talk to us, and even build self-driving cars. Definitely, in the near future, we will witness more serious achievements with this technology, which will allow us to explore new research fields and tasks. The study in this paper belongs to a series of works, where Machine Learning is being used in the study of hard cosmological and astrophysical problems, very difficult to solve otherwise. The approach here is, however, quite different from the ones employed in other works, where observational data have been used to train Machine Learning algorithms. Our approach is a Bayesian Machine Learning algorithm, which allows to discriminate between theories by using model-based forward simulations. In this way, we incorporate our prior believe directly within the posterior distribution. Then, during the Bayesian Learning process, the priors are updated through mock data generation and the Machine Learning algorithm. Recently, Bayesian Machine Learning likelihood-free inference has been successfully applied to study the $H_{0}$ tension problem, by using single viscous fluid models, and also  to constrain the cosmic opacity of the Universe at different redshift ranges~\cite{PyMC3_p1},~\cite{PyMC3_p2}. Having realized the power of the approach through these previous successes, in this paper we have taken a step forward, by incorporating into the analysis string Swampland criteria. More specifically, starting from these criteria and applying then Bayesian Machine Learning, we have been able to constrain the underlying effective theory and to determine if the resulting best theory for describing low redshift behavior of a dark energy dominated Universe will end up in the Swampland or not. 

Our work was motivated by the results in ~\cite{SCGP}, where a Gaussian Process had been used to study string Swampland criteria in a model independent way, directly using observational data without involving a specific dark energy model or the potential describing the scalar field. One of the results from that reconstruction was to realize that an effective theory being in the Swampland could end up or not there. Also, it could start not being  in the Swampland but  end up, again, inside or outside of it. However, that mentioned study  was totally limited to the redshift range were data are already available. On the contrary, using, as we do here, Bayesian Machine Learning, we can overcome these limitations. On the other hand, since the starting point for us is~\cite{SCGP}, we needed one more step in order to perform the generative process based on Bayesian Machine Learning. Specifically, one needs the form of the expansion rate to complete Eqs.~(\ref{eq:SC1}) and~(\ref{eq:SC2}) describing the string Swampland criteria, since they can be expressed in terms of $H(z)$, $H^{\prime}(z)$ and  $H^{\prime\prime}(z)$. Indeed, the consideration of two different models describing $H(z)$ has allowed us to constrain the background dynamics for each of them, and to determine whether the models are in the Swampland or not. It should be mentioned that the results we have obtained are fully consistent with~\cite{SCGP}, what indicates that the string Swampland criterion is in  tension with General Relativity with standard dark matter and quintessence dark energy splitting. Another conclusion is that it should be possible to build an effective theory with phantom dark energy well consistent with the string Swampland criteria. As, moreover, astronomical observations continue supporting the quintessence nature of dark energy,  this could be also seen as an  indication that General Relativity with standard dark matter and quintessence dark energy splitting should eventually end up in the Swampland.

Let us summarize from where we started and what we have achieved up to now. First of all, we have demonstrated that incorporating Bayesian Machine Learning and string Swampland criteria, we can constrain cosmological models. Then, we have validated the  results of the fit by using available expansion rate data. Further, we have seen that the string Swampland criteria and General Relativity with standard dark matter and quintessence dark energy splitting are in some tension. We observed also that, most likely, the theory will not end up in the Swampland, provided we introduce phantom dark energy. All this can be a hint that a simple splitting of the energy source between standard matter and dark energy used in the studies is not applicable on cosmological scales. On the other hand, the results come from a specific form of the expansion rate, Eq.~(\ref{eq:H1}), constructed with the Bayesian Machine Learning approach. It should be mentioned again, that in this procedure one does not use any observational data. In this regard, it becomes interesting to investigate how the form of the expansion rate introduced in the Bayesian Machine Learning approach will affect the Swampland related results. To this end we have studied a second model, where a parameterization of the deceleration parameter  has been used, yielding another parameterization of the expansion rate, Eq.~(\ref{eq:q1_H}). The conclusion is the same, but with one difference: that the model with Eq.~(\ref{eq:q1_H}) is preferable since it can solve the $H_{0}$ tension problem and explain the BOSS result for the $H(z)$ at $z=2.34$. In our analysis, the two redshift ranges $z\in[0,2.5]$ and $z\in[0,5]$ have been considered. Since our results have been validated and compared with the available expansion rate data, we forecast that, most likely, high redshift observations of the expansion rate will not affect  our conclusions on the string Swampland criteria. Moreover, we have used again the Bayesian Machine Learning approach, with a $H(z)$ generative process, to constrain the parameters $H_{0}$ and $H_{1}$ for the first model, Eq.~(\ref{eq:H1}), and $H_{0}$, $q_{1}$ and $q_{2}$  for the second one, Eq.~(\ref{eq:q1_H}). The results indicate that the same approach has imposed very tight constraints on these parameters, as compared with the constraints obtained using $SC1$, Eq.~(\ref{eq:SC1}). This is another evidence of the mentioned existing tension and could be a starting point to finally understanding why it is there and to finding a way to solve it.

To end up,  we would like to remark that an interplay between Bayesian Machine Learning and Swampland criteria can provide a working pipeline allowing to constrain the free parameters of the underlying theory. It can identify the position of the theory in the phase space indicating whether it is in the Swampland or not, and also if eventually it will end up, or not, there. Using two simple models, we have already proven that this approach can be actually useful in this respect, and that it can be used to extract valuable information from a minimal information input. Indeed, there are several  additional questions that should be clarified before we can get a final assessment of the power of the method; but it seems clear that we have initiated a promising path, which has already revealed novel connections between the Swampland and cosmology. We expect to report more results on this topic in forthcoming papers, with particular attention on the construction of a specific form of $H(z)$ exhibiting Swampland-NotSwampland and  NotSwampland-Swampland transitions, as they are not in principle forbidden, according to the study of~\cite{SCGP}. The last point to be mentioned and which will require also future study is the spontaneous sign switch in the dark energy equation of state parameter, which we have observed when the field traverses are in the $z\in[0,5]$ redshift range.

 \section*{Acknowledgement}

This work has been partially supported by MINECO (Spain), project FIS2016-76363-P, and by AGAUR (Catalan Government), project 2017-SGR-247.

\end{document}